\documentclass[aps,prl,twocolumn]{revtex4}
\usepackage{amsmath,graphicx}
\usepackage{pst-all}

\newcommand{\wwr}{|\widetilde{0}\rangle}
\newcommand{\xxr}{|\widetilde{1}\rangle}
\newcommand{\wwl}{\langle\widetilde{0}}
\newcommand{\xxl}{\langle\widetilde{1}}
\newcommand{\square}{p}

\newcommand{\be}{\begin{equation}}
\newcommand{\ee}{\end{equation}}
\newcommand{\bea}{\begin{eqnarray}}
\newcommand{\eea}{\end{eqnarray}}

\begin{document}

\title{Quantum loop models and the non-abelian toric code}

\author{Paul Fendley}
\affiliation{ 
All Souls College and the Rudolf Peierls Centre for Theoretical
Physics,\\ University of Oxford, 1 Keble Road,  OX1 3NP, UK;\\
and Department of Physics, University of Virginia,
Charlottesville, VA 22904-4714 USA
}
 
\date{\today} 
\begin{abstract} 
I define quantum loop models whose degrees of freedom are Ising spins
on the square lattice as in the toric code, but where the excitations
should have non-abelian statistics. The inner product is topological,
allowing a direct implementation of the anyonic fusion matrix on the
lattice. It also makes deconfined anyons possible for a variety of
values of the weight per loop $d$ in the ground state. For
$d=\sqrt{2}$, a gapped non-abelian topological phase can occur with
only four-spin interactions.

\end{abstract} 

\maketitle

Non-abelian anyons have been the subject of intense study recently,
especially because of their potential application to topological
quantum computation \cite{DFNSS}.  It is now well understood how
abelian fractionalized excitations can occur in relatively simple spin
systems, e.g.\ the ``toric code'' \cite{Kitaev97}, and quantum dimer
models \cite{MS}.  An essential property for having deconfined anyons
in such models is that the ground state can be expressed as a
superposition of states comprised of closed loops of all
lengths. Anyons are attached to each other by segments of loop, so
that their braiding is non-trivial when far apart.  There thus has
been considerable effort to find generalizations which have
fractionalized excitations with non-abelian statistics in ``quantum
loop models'' \cite{Freedman01}.

\begin{figure}[t]
\begin{center} 
\includegraphics[height=1.1cm]{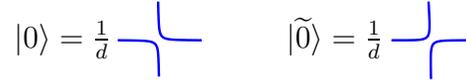}
{\large
    \put(-173,13){$|0\rangle=\frac{1}{d}$}
    \put(-70,13){$\wwr=\frac{1}{d}$}
}
\caption{The two-state quantum system at each vertex}
\label{fig:zerodef}
\end{center}
\end{figure}

The Hilbert space of a quantum loop model is spanned by loop
configurations on some lattice. A Hamiltonian of Rokhsar-Kivelson type
\cite{RK} is chosen so that when the ground state $|\Psi\rangle$ is
written as a sum over different loop configurations, the coefficient
for each loop configuration is the Boltzmann weight of some classical
loop model.  Here I study the ``completely-packed'' loop model, where
every link of the lattice is covered by self- and mutually-avoiding
loops.  There is therefore a quantum two-state system at every vertex,
corresponding to the two ways possible for the loops to avoid each
other \cite{Freedman01,FNS}, shown in fig.\ \ref{fig:zerodef}.
Each loop configuration ${\cal L}$ with $n^{}_{\cal L}$ loops
receives weight $d^{n^{}_{\cal L}}$ in the ground state.
The corresponding classical loop model is the
$Q=d^2$-state Potts model at its self-dual point: the loops surround
clusters in the Fortuin-Kasteleyn expansion \cite{FK}.

When defining a quantum loop model, the inner product as well as the
Hamiltonian must be specified. The simplest inner product is to make
each loop configuration an orthonormal basis element
\cite{Freedman01}. This, however, is undesirable because of the
``$d=\sqrt{2}$'' barrier. Correlators in the ground state of a
quantum-mechanical model are weighted by $|\Psi|^2$. When
each loop configuration is orthonormal to the others,
\begin{equation}
 \langle \Psi| \Psi\rangle= \sum_{\cal L} d^{2n^{}_{\cal L}},
\label{Zloop}
\end{equation}
i.e.\ each loop gets a weight $d^2$. The classical loop partition
function $\langle \Psi|\Psi\rangle$ is dominated by ``short loops''
when the weight per loop $d^2$ is larger than $2$ \cite{Nienhuis}.
This short-loop phase is not critical, so loops of arbitrarily-long
length do not appear in the ground state of the quantum model. This
means that anyons are confined when $d>\sqrt{2}$.

\begin{figure}[h] 
\begin{center} 
\vskip-.02in
\includegraphics[width= .3\textwidth]{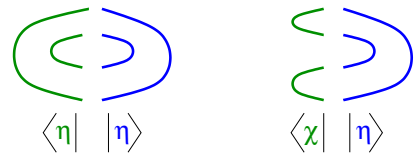} 
\caption{Gluing to find the topological inner product} 
\label{fig:gluing} 
\end{center} 
\vskip-.1in
\end{figure}

There is another reason why the simplest inner product is undesirable
\cite{Levin}. Non-abelian anyons have loop segments attached to them.  Consider
two such states with anyons at the
same locations. One then can ``glue'' the dangling loop ends of the two
together, so that the combined
configuration consists entirely of closed loops. To obtain a
topological theory in three-dimensional spacetime, each loop formed by
this gluing must contribute a factor of $d$ to the topological inner
product, just as loops in the ground state do. For example, let
$|\eta\rangle$ and $|\chi\rangle$ each have four anyons in the same
places, but let the loop ends be connected in different ways.  Computing
the inner product is easily done via the schematic pictures in
figure \ref{fig:gluing}, giving
$\langle \eta|\eta\rangle=\langle
\chi|\chi\rangle=d^2$, while $\langle \chi|\eta\rangle=d$. Loop
configurations are not orthonormal with the
topological inner product.

In this paper I define the completely-packed quantum loop model on the
square lattice with a local and topological inner product.  With this
inner product, the partition function $\langle \Psi| \Psi\rangle$ of
the corresponding classical model is no longer given by (\ref{Zloop}),
but by a different classical loop model. There is good evidence that
the loops in the new classical model are critical for values of $d$
larger than $\sqrt{2}$, so it is possible to kill two birds with one
inner product.

The topological inner product in the completely-packed quantum loop
model is found by considering the entire system to be a single vertex,
so that $|0\rangle$ and $\wwr$ are the only two states in the
system. Each has four ends, and the topological inner product is
computed by gluing ends at the same site together. Gluing $\langle 0|$
to $|0\rangle$ gives two loops, as in the left of figure
\ref{fig:gluing}. Each loop contributes a factor $d$, so the $1/d$ in
front in figure \ref{fig:zerodef} normalizes $\langle 0|0\rangle =
\wwl\wwr=1$. Gluing $\langle 0|$ to $\wwr$ gives a single loop, as in
the right of figure \ref{fig:gluing}. The demand of topological
invariance therefore requires that at every vertex
\begin{equation}
\langle 0|0\rangle = \wwl\wwr=1, 
\qquad \langle 0\wwr=1/d.
\label{topip}
\end{equation}
This inner product is positive definite when $|d|>1$;
note that $d$ can be negative.

\begin{figure}[h] 
\begin{center} 
\includegraphics[width= .3\textwidth]{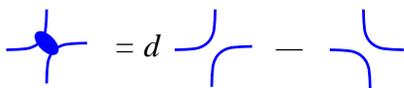} 
\caption{$|1\rangle$ (up to overall normalization) in terms of loops;
  $\xxr$ has each term rotated by 90 degrees.}
\label{fig:spin1} 
\end{center}
\end{figure}

This choice of inner product is very
natural. As illustrated in figure
\ref{fig:spin1}, define
\begin{eqnarray}
\label{def1}
|1\rangle &=& \frac{1}{\sqrt{d^2-1}}\left(d\wwr-|0\rangle\right)\\
|\widetilde{1}\rangle &=& 
\frac{1}{\sqrt{d^2-1}}\left(d|0\rangle-\wwr\right).
\label{def1tilde}
\end{eqnarray}
This yields
$\langle 0|1\rangle =\wwl\xxr=0$ and $\langle 1|1\rangle
=\xxl\xxr=1.$ 
There are therefore two natural orthonormal bases for the Hilbert space
at each site, one with basis elements $|0\rangle,|1\rangle$, 
and the other with basis elements $\wwr,\xxr$. The unitary
transformation relating them is
\begin{equation}
\begin{pmatrix}
\wwr\\
\xxr
\end{pmatrix}
= F 
\begin{pmatrix}
|0\rangle\\
|1\rangle
\end{pmatrix}
;\quad
F=\frac{1}{d}
\begin{pmatrix}
1 & \sqrt{d^2-1}\\
\sqrt{d^2-1} & -1
\end{pmatrix}\ .
\label{Fdef}
\end{equation}
This is precisely the desired fusion matrix for non-abelian anyons in
the quantum loop model, or equivalently, the fusion matrix of the
conformal field theory $SU(2)_k$ with $d=2\cos(\pi/(k+2))$ for $k$
integer.

The reason why this is exactly what we want goes to the very heart of
the quantum loop model. Excitations in the quantum loop model are
obtained by cutting a loop, so that two quasiparticles are attached by
a zero-energy strand. The weighting of $d$ per loop in the ground
state is necessary for the excitations to have consistent non-abelian
braiding and fusing (in mathematical language, these are the
conditions for a conformal field theory/modular tensor category)
\cite{DFNSS}. Namely, in the $SU(2)_k$ theories considered here, these
consistency conditions require that each type of anyon corresponds to
a representation of the quantum-group algebra $U_q(sl(2))$. One
doesn't need to know much about this algebra other than that it has
spin-$s$ representations like ordinary $su(2)$ for $s\le k/2$, and
that their tensor product behaves similarly that of $su(2)$, once one
takes this truncation into account. In particular for $k\ge 2$, two
spin-1/2 representations fuse to either spin 0 or spin 1. With these
rules, like with $su(2)$, there are two linearly-independent ways that
four spin-1/2 representations can fuse to spin 0.



The anyonic structure is beautifully realized in this
quantum loop model. Since anyons are attached to strands, the strand
must fuse as if they have spin 1/2. Two anyons
connected by a strand fuse to the identity: the combination has
trivial statistics. Demanding that the strands form closed loops in
the ground state thus requires that the four strands at
each vertex on the square lattice fuse to the identity. 
Labeling the four strands
at a vertex (or equivalently, the four spin-1/2 representations) by
$a,b,c,d$, the state $|0\rangle$ corresponds to requiring that the
pair $a,b$ fuses to the identity (spin 0). Since all four strands fuse
to the identity, the pair $c,d$ must then also fuse to the identity
($0\times 0=0$). Likewise, the state $\wwr$ corresponds to the
pair $b,c$ and the pair $a,d$ each fusing to the identity. As if these
were $su(2)$ spins, these two states are not orthogonal. Instead, the
state orthogonal to $|0\rangle$ corresponds to the pair $a,b$ and the
pair $c,d$ each fusing to spin 1, which is denoted as the state
$|1\rangle$. The basis $\wwr,\xxr$ is defined analogously using the
pairs $a,d$ and $b,c$. The change-of-basis matrix $F$ in the fusion
algebra is the same as that obtained in (\ref{Fdef}).  Thus the
topological inner product indeed requires that the strands 
behave as the spin-1/2 representation of $SU(2)_k$.

This Hilbert space and ground state are related to those of
``string-net'' models \cite{LevinWen}, where non-abelian anyons arise
by fine-tuning the Hamiltonian to reflect the fusion matrix of the
desired topological field theory.
For the $SU(2)_k$ string-net model, the states allowed on each link of
the honeycomb lattice are labeled by the representations of spin
$j=0,1/2,\dots, k/2$.  Analogously to \cite{Freedman01} and here, the
ground state is comprised solely of states where the
representations on the links touching each vertex all fuse to the
identity.
The Hilbert space of this paper is obtained by 
restricting the $SU(2)_k$ string-net model so
that {\em all} the links in two of the three directions of the
honeycomb lattice have $j=1/2$, because requiring that every trivalent vertex
fuse to the identity means that the states on the remaining links can
take the states $j=0$ or $1$. They correspond to the states
$|0\rangle$ and $|1\rangle$ on the square lattice simply by
compressing each such link to a point.  The $\wwr,\xxr$ basis at a
vertex corresponds to stretching out each square vertex in the
orthogonal direction. 

\begin{figure}[h] 
\begin{center} 
\includegraphics[width= .35\textwidth]{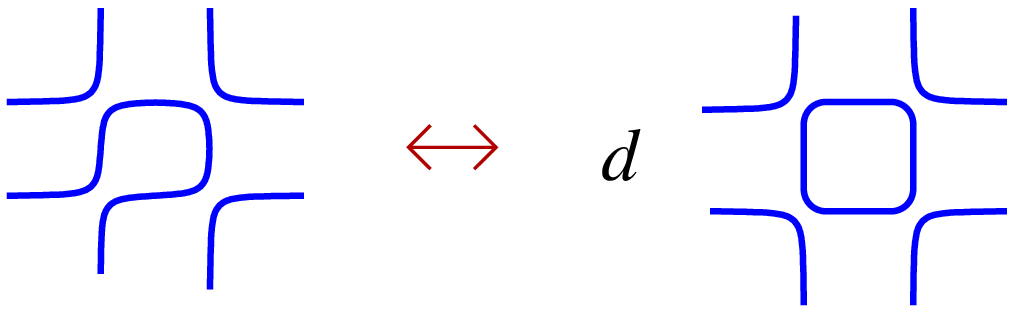} 
\caption{The flip term ${\cal F}_0P_{\,\widetilde{0}}^{}P_0^{} P_{\,\widetilde{0}}^{}$} 
\label{fig:flip} 
\end{center} 
\end{figure}

Now that the Hilbert space and the inner product have been specified,
I construct a Rokhsar-Kivelson-type Hamiltonian (i.e.\
``rokk'' the Hilbert space) whose ground state is a sum over loop
configurations with weight $d$ per loop. This Hamiltonian is a sum
over projection operators which annihilate the ground state, and all
eigenstates with non-zero eigenvalues are orthogonal to the ground
state. The off-diagonal ``flip'' part acts on
plaquettes with all four links belonging to the same loop, as
illustrated in figure \ref{fig:flip}.
On the plane, one can connect any configuration to any other by doing
flips like this and its rotations by 90 degrees \cite{Freedman01,FNS}. 

The projection operators which implement the flip and
enforce the weight $d$ per loop in the ground state
therefore have off-diagonal terms flipping between $|0\rangle$ and $\wwr$.
In the $|0\rangle,\wwr$ basis, these operators are
\begin{equation}
\widehat{\cal F}_0^{}\propto
\begin{pmatrix}
\frac{1}{d}&-1\\
-1&d
\end{pmatrix}
,\qquad
\widehat{\cal F}^{}_{\,\widetilde 0}
\propto
\begin{pmatrix}
d&-1\\
-1&\frac{1}{d}
\end{pmatrix},
\label{FFdef}
\end{equation}
where the subscript indicates the configuration with larger weight in
the ground state. Operators in the $|0\rangle,\wwr$ basis
have a hat and those in the $\wwr,\xxr$ basis have a tilde. 
The projectors in the orthonormal bases are
found from (\ref{FFdef}) by the
non-unitary change of basis
$$
\begin{pmatrix}
|0\rangle\\
\wwr
\end{pmatrix}
= V 
\begin{pmatrix}
|0\rangle\\
|1\rangle
\end{pmatrix}
,\qquad
V=\frac{1}{\sqrt{d^2-1}}
\begin{pmatrix}
\sqrt{d^2-1} & -1\\
0 & d
\end{pmatrix}\ .
$$
In the orthonormal $|0\rangle,|1\rangle$ basis the flips are therefore
$${\cal F}_0^{}=
\begin{pmatrix}
\frac{\sqrt{d^2-1}}{d^2+1}&-1\\
-1&\frac{d^2+1}{\sqrt{d^2-1}}
\end{pmatrix}
,\qquad
{\cal F}^{}_{\,\widetilde 0}
=
\begin{pmatrix}
\frac{\sqrt{d^2-1}}{2}&-1\\
-1&\frac{2}{\sqrt{d^2-1}}
\end{pmatrix}\ .
$$

A slightly confusing fact is that the operator which
projects onto $|0\rangle$ in the $|0\rangle,\wwr$ basis is not 
diagonal when transformed to the $|0\rangle, |1\rangle$ basis. 
Rather, 
$$P_0^{} \equiv V^T \widehat{P}_0^{} V=
\begin{pmatrix}
{\sqrt{d^2-1}}&-1\\
-1&\frac{1}{\sqrt{d^2-1}}
\end{pmatrix} \ .
$$ ($\widehat{P}_0^{}\propto\begin{pmatrix}1&0\cr 0&0\end{pmatrix}$ is
normalized so that $P_0^{}$ is a projector.)  
The
off-diagonal pieces are a consequence of $\wwl|0\rangle\ne 0$: $P_0^{}$
should be understood as projecting onto a state orthogonal to $\wwr$,
i.e.\ in the $\wwr,\xxr$ basis
$$\widetilde{P}_0^{}= FP_0^{}F=  \widetilde{P}^{}_{\,\widetilde{1}} \ .$$ 
Likewise, the operator $P^{}_{\,\widetilde 0}$ projects onto a state
orthogonal to $|0\rangle$, i.e.\ $|1\rangle$: 
$$P^{}_{\,\widetilde 0} = V^T \widehat{P}^{}_{\,\widetilde 0} V \ .
=\begin{pmatrix}
0&0\\
0&1
\end{pmatrix}\ =P^{}_1\ .
$$


The Hamiltonian acts on a two-state ``spin'' system
$|0\rangle,|1\rangle$ at each site of the lattice, each term involving
four spins around a plaquette. Then $H=\sum_{\square} {\cal
F}_{\square}$, with
$$
{\cal F}_{\square}
={\cal F}_0^{} P^{}_{\,\widetilde 0} P_0^{} P^{}_{\,\widetilde 0}
+P_0^{} {\cal F}^{}_{\,\widetilde 0} P_0^{} P^{}_{\,\widetilde 0}
+P_0^{} P^{}_{\,\widetilde 0} {\cal F}_0^{} P^{}_{\,\widetilde 0}
+P_0^{} P^{}_{\,\widetilde 0} P_0^{} {\cal F}^{}_{\,\widetilde 0}
$$
where the first projector in each term acts on the lower-left spin on
the plaquette, the second on the upper-left, and so on clockwise.
By construction, the sum over all configurations with weight $d$ per
loop is annihilated by $H$. Since this Hamiltonian is the sum of
projection operators, all eigenvalues must be greater than equal to
zero. When space is a plane, repeated
applications of $H$ connect all the configurations, so the ground
state is unique.

\begin{figure}[h] 
\begin{center} 
\includegraphics[width= .1\textwidth]{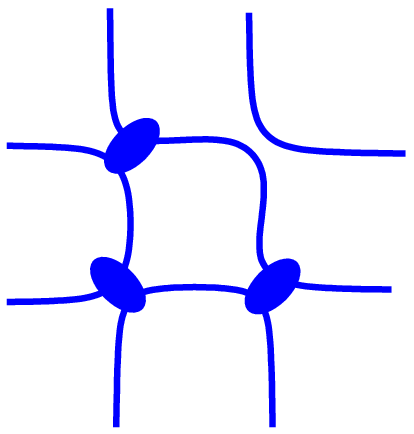} 
\caption{The Jones-Wenzl projector ${\cal F}_0^{} {\cal F}^{}_{\,\widetilde 0}
P_0^{}  {\cal F}^{}_{\,\widetilde 0}$
for $d=\pm\sqrt{2}$} 
\label{fig:JW2} 
\end{center} 
\end{figure}

On the torus, however, the ground state is not unique, because ${\cal
F}_\square$ will not change the number of loops which are wrapped
around a cycle of the torus -- it only creates and annihilates loops
locally. To get a finite number of ground states on the torus, one
must therefore add another term to the Hamiltonian. A local term which
breaks this degeneracy while still preserving the weight $d$ per loop
in the ground state is possible only when
$d=2\cos(\pi j/(k+2))$ with $j$ and $k+2$ coprime integers
\cite{Freedman01}.
This term is known as the Jones-Wenzl projector, and in the algebraic
picture, it projects onto the state of spin $(k+1)/2$,
whose corresponding anyon is not part of the spectrum.  For
$d=\sqrt{2}$, fusing two spin-1 anyons gives only the identity sector
(only the spin-1/2 anyon is non-abelian when $k=2$). Therefore, fusing
three spin-1 anyons here cannot give the identity: $1\otimes (1
\otimes 1)= 1\otimes 0 = 1$. Since the topological characteristics of
the ground state should reflect this fusion algebra, loop
configurations involving the three spin-1 strands in figure
\ref{fig:JW2} should be forbidden from the ground state.

The projector onto the state pictured in figure
\ref{fig:JW2} is nice in the square-lattice model here, since it
involves only the spins around a single plaquette.  It involves only
projectors given above, because the flip operators
$\widehat{\cal F}^{}_0$ and $\widehat{\cal F}^{}_{\,\widetilde 0}$
introduced in (\ref{FFdef}) project onto $|1\rangle$ and $\xxr$
respectively: $|1\rangle$ is the eigenstate of the
projection operator $\widehat{\cal F}_0$ with non-vanishing
eigenvalue. Therefore, the Jones-Wenzl projector for a plaquette in
the $|0\rangle,|1\rangle$ basis at $k=2$ is
$$
{\cal J}_\square
={\cal F}_0^{} {\cal F}^{}_{\,\widetilde 0} {\cal F}_0^{} P^{}_{\,\widetilde 0}
+{\cal F}_0^{} {\cal F}^{}_{\,\widetilde 0} P_0^{} {\cal F}^{}_{\,\widetilde 0}
+{\cal F}_0^{} P^{}_{\,\widetilde 0} {\cal F}_0^{}  {\cal F}^{}_{\,\widetilde 0}
+P_0^{} {\cal F}^{}_{\,\widetilde 0} {\cal F}_0^{} {\cal F}^{}_{\,\widetilde 0}
$$ 
where the first projector in each term acts on the lower-left spin
as before. Adding this to the Hamiltonian ensures that the combination of loops illustrated in figure (\ref{fig:JW2}) is orthogonal to the ground state.

The full Hamiltonian for $d=\sqrt{2}$, with nine
ground states on the torus, is then given by
$H=\sum_\square \left[{\cal F}_\square + {\cal J}_\square\right].$ It
can be written out entirely in terms of Pauli matrices $\sigma^x$ and
$\sigma^z$ acting at each site, with each term involving four Pauli
matrices acting around a plaquette. Applying the $F$ matrix
(\ref{Fdef}) at $d=\sqrt{2}$ means that changing between 
$|0\rangle,|1\rangle$ and $\wwr,\xxr$ bases amounts to
interchanging $\sigma^x$ and $\sigma^z$ in the Hamiltonian.
This resembles the toric code \cite{Kitaev97} if one treats each spin as
living on the {\em links} of another square lattice with unit cell of
twice the area as the original lattice. Then the Hamiltonian divides
into terms acting on the four links around each site, and the four
links around plaquette of this new lattice.


One way of obtaining non-abelian anyons in the spectrum is by
allowing empty links (defects) into the model.
However, expanding the Hilbert space is probably unnecessary.
Because ${\cal F}_0$ projects onto spin $1$, the net spin of a
plaquette of spins not annihilated by ${\cal F}_\square$ is also 1.
For $k>2$ spin-1 anyons have non-abelian statistics, but even for
$k=2$, these excitations can braid non-trivially because of the two
spin-1/2 strands attached to each.  It is not yet proven whether or
not such excitations are gapped (the proof of gaplessness in
\cite{FNS} does not apply because of the different inner product). It
seems very likely that once the Jones-Wenzl projector is included,
they will be gapped. Each ${\cal J}_p$ can be added to the Hamiltonian
with any coefficient without changing the ground state, so for an
excited state not to be gapped, it would need to be annihilated
by all ${\cal J}_p$ like the ground state.

With the topological inner product, it is possible to crack the
$d=\sqrt{2}$ barrier. To have deconfined anyons, the corresponding
classical loop model must be critical \cite{Freedman01, FNS}. The
classical partition function is no longer simply (\ref{Zloop}), but
instead is given by a sum over two completely-packed loop configurations
${\cal L}$ and ${\cal L}'$ on the same lattice:
\begin{equation}
\langle \Psi|\Psi\rangle 
= \sum_{{\cal L},{\cal L}'} d^{n^{}_{\cal L}}
  d^{n_{{\cal L}'}} \lambda^{n_X^{}}
\label{Zloop2}
\end{equation}
where $n_X^{}$ is the number of vertices at which ${\cal L}$ and ${\cal
L}'$ differ.  With the topological inner product, $\lambda=1/d$, while
(\ref{Zloop}) from the simple inner product is recovered when
$\lambda\to 0$. Since taking the inner product of a given
configuration with itself always gives an even power of $d$, changing
$d\to -d$ is equivalent to instead flipping $\lambda\to -\lambda$.
(Note that in the orthonormal bases the Hamiltonian depends only on $d^2$.)
Since the partition function of the $Q$-state Potts model can be
expanded in terms of completely-packed loops with weight $\sqrt{Q}$
each, (\ref{Zloop2}) describes two $d^2$-state Potts models coupled by
a self-dual perturbation. When $\lambda=\pm 1$, the models are
decoupled. For $\lambda\to 0$ one obtains a single $Q=d^4$ state Potts
model at its self-dual point, which indeed is critical only for $d\le
\sqrt{2}$.  

When $d=\sqrt{2}$, the partition function (\ref{Zloop2}) is the same
as that for the Ashkin-Teller model, i.e.\ two coupled Ising
models. For any 
$\lambda\ge 0$, including $\lambda=1/d=1/\sqrt{2}$, 
the model remains critical \cite{Nienhuis}. Along this critical line,
the energy operator (odd under duality) is of dimension $x=2/g$,
where $g= 8/\pi \sin^{-1}[1/2 + 1/(2+2\sqrt{2}\lambda)]$. However,
understanding the fractal properties of the loops as a function of
$\lambda$ seems to be an open problem.


As opposed to the classical loop model with partition function
(\ref{Zloop}), the model with (\ref{Zloop2}) can be critical even for
$|d|=2\cos(\pi/(k+2))>\sqrt{2}$ if $\lambda$ is {\em negative}. By using
level-rank duality of the loop representation of the BMW algebra
\cite{FKrush}, it is shown in \cite{FJ} that there is a critical point
when $\lambda=\lambda_c$, where
$\lambda_c=-\sqrt{2}\sin(\pi(k-2)/[4(k+2)])$. Moreover, numerical
evidence strongly suggests that for $1/\lambda_c<\lambda<\lambda_c$, the
classical loop model has a critical phase \cite{FJ}. When $k=6$,
$\lambda_c=-1/d$. Thus for $k<6$, the classical loop model at
$\lambda=-1/d$ falls into this phase, deconfining anyons in
the quantum model.


This quantum loop model can be generalized by 
relaxing the requirement that all links be in the same representation
\cite{Freedman01,LevinWen}, or by studying different lattices. The
completely-packed model on the square lattice discussed here
simplifies in several very nice ways, e.g.\ at $d=\sqrt{2}$ the
Jones-Wenzl projector is no more complicated than the flip term. On
the Kagom\'e lattice at $d=\sqrt{2}$ it should be even simpler, only
involving three-spin interactions. It would be very interesting to
find other lattices and other representations with such elegant
properties.


\medskip

I am grateful to Michael Freedman for explaining Michael Levin's
observation to me, and for many conversations on quantum loop models.
This research has been supported by the NSF under grants DMR-0412956
and DMR/MSPA-0704666, and by an EPSRC grant EP/F008880/1.


\def\cmp#1#2#3{Comm.\ Math.\ Phys.\ {\bf #1}, #2 (#3)}

\end{document}